\documentclass[twocolumn,nofootinbib,amsmath,amssymb,a4paper]{revtex4}

\usepackage{graphicx}
\usepackage{dcolumn}
\usepackage{bm}

\newcommand{\be}{\begin{equation}}
\newcommand{\ee}{\end{equation}}
\newcommand{\bea}{\begin{eqnarray}}
\newcommand{\eea}{\end{eqnarray}}
\newcommand{\nn}{\nonumber}

\newcommand{\DB}{\Delta B}

\begin{document}

\title{B-Hadron Lifetimes, Width Differences and Semileptonic CP-Asymmetries}

\author{Cecilia Tarantino}
 \email{cecilia.tarantino@ph.tum.de}
\affiliation{
Physik Department,\\
Technische Universit\"at M\"unchen,\\
D-85748 Garching, Germany}

\begin{abstract}
$B$-hadron physics plays a fundamental role to test and improve our
understanding of flavor dynamics within and beyond the Standard Model.
Of particular phenomenological interest are beauty hadron
lifetime ratios and, width differences and semileptonic CP-asymmetries in 
$B_d$ and $B_s$ systems.
We discuss their theoretical predictions which, in the last years, have
been improved thanks to accurate lattice calculations of operator matrix
elements and 
perturbative computations of Wilson coefficients.
\end{abstract}

\maketitle

\section{Introduction}
\label{intro}
Neutral $B_d$ and $B_s$ mesons mix with their antiparticles leading to 
oscillations between the mass eigenstates. The time evolution of the neutral 
meson doublet is described by a Schr\"odinger equation with an effective
$2 \times 2$ Hamiltonian
\bea
i\frac{d}{dt} \left(\hspace*{-0.1cm}\begin{array}{c} B_q\\ {\overline B}_q \end{array} \hspace*{-0.1cm}\right)
\hspace*{-0.1cm}=\hspace*{-0.1cm}
\left[ \left(\hspace*{-0.1cm}\begin{array}{cc} M_{11}^q & M_{12}^q\\ {M_{12}^{q}}^* & M_{11}^q
\end{array}\hspace*{-0.1cm}\right)\hspace*{-0.1cm} 
-\hspace*{-0.1cm}\frac{i}{2}\left(\hspace*{-0.1cm}\begin{array}{cc}\Gamma_{11}^q & \Gamma_{12}^q\\
{\Gamma_{12}^{q}}^* & \Gamma_{11}^q
\end{array}\hspace*{-0.1cm}\right) \right]\hspace*{-0.1cm}
\left(\hspace*{-0.1cm}\begin{array}{c} B_q\\ {\overline B}_q \end{array} \hspace*{-0.1cm}\right).
\label{eq:schro}
\eea
Mass and width differences are defined as 
$\Delta m_q=m^q_H-m^q_L$ and $\Delta\Gamma_q=\Gamma^q_L-\Gamma^q_H$,
where $H$ and $L$ denote the Hamiltonian eigenstates with the heaviest and
lightest mass eigenvalue. 
These states can be written as
\be
\vert B_q^{L,H}\rangle ={1\over \sqrt{1+\vert (q/p)_q \vert^2}}\,\left(
\vert B_q\rangle \pm  \left(q/p\right)_q\vert {\overline B}_q\rangle
\right).
\ee

The theoretical expressions of hadron lifetimes are related to $\Gamma^q_{11}$ 
($\tau(B_q)=1/\Gamma^q_{11}$), while width differences 
$\Delta \Gamma_q$ and semileptonic CP-asymmetries $A_{SL}^q$ are related to $M^q_{12}$ and 
$\Gamma^q_{12}$.
In $B_{d,s}$ systems, the ratio $\Gamma^q_{12}/M^q_{12}$ is of
${\cal O}(m_b^2/m_t^2)\simeq 10^{-3}$. Therefore, by neglecting terms of ${\cal 
O}(m_b^4/m_t^4)$, one can write
\be
\Delta \Gamma_q=-2\,\vert M^q_{12}\vert\,
{\mathrm{Re}}\left(\frac{\Gamma^q_{12}}{M^{q}_{12}}\right)\,,\qquad
A_{SL}^q= {\mathrm{Im}}\left(
\frac{\Gamma^q_{12}}{M^q_{12}}\right)\,.
\label{eq:dgammared}
\ee

The matrix elements $M^q_{12}$ and $\Gamma^q_{12}$ are related, respectively, 
to the dispersive and the absorptive parts of $\DB=2$ transitions. 
In the Standard Model (SM), these transitions are the result of second-order charged weak 
interactions involving the well-known box diagrams.

The dispersive matrix element $M^q_{12}$ has been computed at 
the NLO in QCD~\cite{Buras:1990fn}.
In presence of New Physics (NP) $M^q_{12}$ is modified by the contribution of
new heavy particles running in the loop.
Both $\Delta \Gamma_q$ and $A_{SL}^q$, therefore, result to be
of great phenomenological interest for their sensitivity to 
NP~\cite{Grossman:1996er,Dunietz:2000cr,Laplace:2002ik,UTfit,BBGT,Ligeti:2006pm}.
 
The absorptive matrix elements $\Gamma^q_{11}$  and 
$\Gamma^q_{12}$ can be computed by applying the heavy quark expansion 
(HQE)~\cite{ope}, with a consequent separation of
short-distance from long-distance contributions.
The great energy ($\sim m_b$) released in beauty hadron decays, in fact,
allows  to expand the inclusive widths in powers of $1/m_b$.
Theoretical predictions of inclusive rates are based on a 
non-perturbative calculation of matrix elements, mainly studied in lattice QCD,
 and a perturbative calculation of Wilson coefficients.

The contributions of light quarks in beauty hadron decay widths 
(spectator effects) have been computed at $\mathcal{O}(\alpha_s)$ in QCD and 
$\mathcal{O}(\Lambda_{QCD}/m_b)$ in the HQE.
Based on these calculations are the theoretical predictions for beauty 
hadron lifetimes and B-meson width differences and semileptonic
CP-asymmetries.  
Improved theoretical estimates have been
obtained, to be compared with recent experimental measurements or
limits.


\section{Beauty hadron lifetime ratios}
\label{sec:1}

The experimental averages of the lifetime ratios of beauty hadrons 
are~\cite{BBpage}
\bea
\label{eq:rexp}
&\frac{\tau(B^+)}{\tau(B_d)}=1.076 \pm 0.008 \,, 
\frac{\tau(B_s)}{\tau(B_d)}=0.957 \pm 0.027 \,,& \nn\\
&\frac{\tau(\Lambda_b)}{\tau(B_d)}=0.84 \pm 0.05\,.&
\eea

By applying the HQE, the inclusive decay width of a hadron $H_b$ can be
expressed as a sum of contributions of local $\DB=0$ operators with increasing dimension, as
\be
\qquad\Gamma (H_b) = \sum_k \frac{\vec{c}_k (\mu)}{m_b^k}\,\langle H_b | 
\vec{O}^{\DB = 0}_k (\mu) | H_b \rangle\,.
\label{eq:sum}
\ee
The HQE yields the separation of short distance effects, confined in Wilson 
coefficients ($\vec{c}_k$), from long distance physics, represented by 
matrix elements of the local operators ($\vec{O}_k^{\DB=0}$).

Spectator contributions, which distinguish different beauty 
hadrons, appear at $\mathcal{O}(1/m_b^3)$  in the HQE.
These effects, although suppressed by an additional power of $1/m_b$, are 
enhanced with respect to the leading contributions by a phase-space factor of 
$16 \pi^2$, being $2 \to 2$ processes instead of $1 \to 3$ 
decays~\cite{Bigi:1992su,NS}.
In order to evaluate the spectator effects one has to calculate the matrix 
elements of dimension-six current-current and penguin operators, 
non-perturbatively, and their Wilson coefficients, in perturbation theory.

Concerning the perturbative part, NLO QCD corrections to the coefficient
functions of the current-current operators have been
computed~\cite{Ciuchini:2001vx}-\cite{NOI}.

Concerning the non-perturbative part, the non-valence contributions,
corresponding to contractions of two light quarks in the same point,
have not been computed. Their non-perturbative lattice calculation would be 
possible, in principle, however it requires to deal with the difficult problem 
of power-divergence subtractions.
On the other hand, the valence contributions, which exist when the light quark
of the operator enters as a valence quark in the external hadronic state, have 
been evaluated. 
For $B-$mesons, several (quenched) lattice studies~\cite{Becirevic:2001fy}-\cite{DiPierro:1998cj} in QCD, HQET and NRQCD exist and yield
compatible results, while for the 
$\Lambda_b$ baryon, only one (quenched) lattice calculation in HQET has been
performed~\cite{DiPierro:1999tb} so far.

More recently, the sub-leading spectator effects which appear at 
$\mathcal{O}(1/m_b^4)$ in the HQE, have been included in the analysis of 
lifetime ratios.
The relevant operator matrix elements have been estimated in the vacuum 
saturation approximation (VSA) for $B-$mesons and in the quark-diquark model 
for the $\Lambda_b$ baryon, while the corresponding Wilson coefficients have 
been calculated at leading order (LO) in QCD~\cite{Gabbiani:2003pq}.

\begin{figure}
\includegraphics[width=0.4\textwidth]{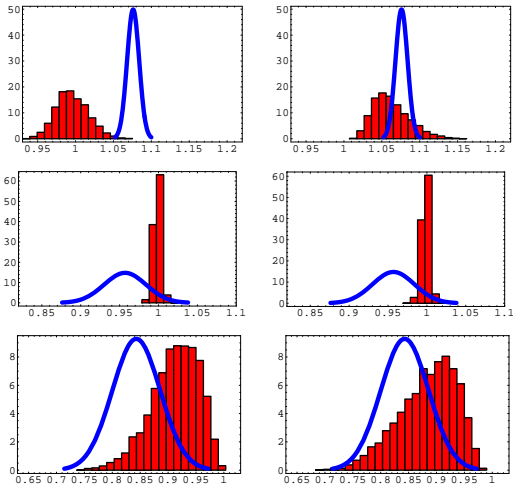}
\vspace*{-0.4cm}
\caption{\label{fig:plot} Theoretical (histogram) vs experimental (solid line) 
distributions of lifetime ratios. The theoretical predictions are shown 
at the LO (left) and NLO (right).}
\end{figure}

The theoretical predictions for the lifetime ratios read~\cite{Tarantino}
\bea
\label{eq:nlores}
&\frac{\tau(B^+)}{\tau(B_d)} =
1.06 \pm 0.02 \,,\qquad 
\frac{\tau(B_s)}{\tau(B_d)}  =
1.00 \pm 0.01 \,,&  \nn\\
& \frac{\tau(\Lambda_b)}{\tau(B_d)}  =
0.88 \pm 0.05\,.&
\eea
They turn out to be in good agreement with the experimental measurements of 
Eq.~(\ref{eq:rexp}).

It is worth noting that the agreement at $1.2 \sigma$ between the theoretical
prediction for $\tau(\Lambda_b)/\tau(B_d)$ and its experimental value
is achieved thanks to the inclusion of NLO (see Fig.~\ref{fig:plot}) and
$1/m_b$ corrections to spectator effects. 
They both decrease the central value of $\tau(\Lambda_b)/\tau(B_d)$ by $8$\%
and $2$\% respectively.

Few months ago, however, the experimental picture was modified by new Tevatron
measurements~\cite{lambdaCDF,lambdaD0}
\be
\frac{\tau(\Lambda_b)}{\tau(B_d)}=1.041(57)\,[\text{CDF}]\,,\,\,
\frac{\tau(\Lambda_b)}{\tau(B_d)}=0.870(102)(41)\,[\text{D0}]\,.
\ee
The uncertainty of the preliminary D0 result is still too large for a
significant comparison, while the CDF value is
surprisingly higher ($\sim 2.5 \sigma$) than the world average.
Although the CDF result represents the single best measurement from fully
reconstructed $\Lambda_b \rightarrow \Lambda^0 J_\psi$ decay, one has to wait
for experimental updates before interpreting the difference of the CDF result
from theory as a $\Lambda_b$-puzzle (reversed w.r.t to the old one, as now
the measurement is larger than the theoretical prediction).

On the theoretical side, further improvement of the $\tau(\Lambda_b)/\tau(B_d)$ theoretical prediction 
would require the calculation of the current-current operator non-valence 
B-parameters and of perturbative and non-perturbative contributions of the 
penguin operator, which appears at NLO and whose matrix elements present
the same problem of power-divergence subtraction.
These contributions are missing also in the theoretical predictions of 
$\tau(B^+)/\tau(B_d)$ and $\tau(B_s)/\tau(B_d)$, but in these cases they 
represent an effect of $SU(2)$ and $SU(3)$ breaking respectively, and are 
expected to be small.

\section{Width differences}
\label{sec:2}
The width differences between ``light''  and ``heavy'' neutral $B_q$-mesons 
($q=d, s$) are defined in terms of the off-diagonal matrix elements 
$\Gamma^q_{12}$ (see Eq.~(\ref{eq:dgammared})).

In the HQE of $\Gamma^q_{12}$, the leading contributions come at 
$\mathcal{O}(1/m_b^3)$ and are given by dimension-six $\DB=2$ operators.
Up to and including $\mathcal{O}(1/m_b^4)$ contribution, one can write 
\bea
\Gamma^q_{12} =
-\frac{G_F^2 m_b^2}{24 \pi M_{B_q}}\cdot
\left[
c^q_1(\mu_2) {\langle  B_q \vert {\cal O}^q_1(\mu_2) \vert \overline B_q\rangle}
+ \right.\nn\\
\left.c^q_2(\mu_2) {\langle B_q \vert {\cal O}^q_2(\mu_2) \vert \overline B_q\rangle} 
+ \delta^q_{1/m}\right]\, ,
\label{eq:gamma12q}
\eea
where ${\langle \overline B_q \vert {\cal O}^q_i(\mu_2) \vert B_q\rangle}$ are the 
matrix elements of the two independent dimension-six operators
\bea
{\cal O}^q_1 &=& \bar b_i \gamma^\mu (1 - \gamma_5)
  q_i\,\bar b_j \gamma^\mu (1 - \gamma_5) q_j\,,\nn\\ 
{\cal O}^q_2 &=& \bar b_i (1 - \gamma_5)
  q_i\,\bar b_j (1 - \gamma_5) q_j\,,
\label{eq:oldbasis}
\eea
 $c^q_i(\mu_2)$
their Wilson coefficients, known at the NLO in 
QCD~\cite{BBlargh}-\cite{Beneke:2003az}, while $\hat{\delta}_{1 / m_b}$ 
represents the contribution of dimension-seven operators~\cite{seven}.

Lattice results of the dimension-six operator matrix 
elements~\cite{Gimenez:2000jj}-\cite{Lellouch:2000tw} have been confirmed and 
improved, by combining QCD and HQET results in the heavy quark 
extrapolation~\cite{damir}.
Moreover, the effect of including dynamical quarks has been examined,
within the NRQCD approach, finding that these matrix elements are essentially 
insensitive to switching from $n_f=0$ to 
$n_f=2$~\cite{Yamada:2001xp,Aoki:2003xb} and to $n_f=2+1$~\cite{staggered}.

Concerning the dimension-seven operators, their matrix elements have never
been  estimated out of the VSA.
Two of these four matrix elements, however, can be related through Fierz 
identities and equations of motion to the complete set of operators studied in~\cite{damir}.
For the other two, a QCD-sum rule calculation is in progress~\cite{pivovarov}.

The theoretical predictions given in~\cite{NOIwip} read
\be
\Delta \Gamma_{d}/\Gamma_{d} = (2.3 \pm 0.8)\cdot 10^{-3}\,,\quad
\Delta \Gamma_{s}/\Gamma_{s} = (7 \pm 3)\cdot 10^{-2} \,,
\label{eq:dg_noi}
\ee
compatible with the experimental averages~\cite{BBpage,FPCP}
\be
\Delta \Gamma_{d}/\Gamma_{d} = (9 \pm 37) \cdot 10^{-3} \,,\quad
\Delta \Gamma_{s}/\Gamma_{s} = (14 \pm 6) \cdot 10^{-2}\,,
\ee
within  quite large uncertainties.

\begin{figure}
\includegraphics[width=0.5\textwidth]{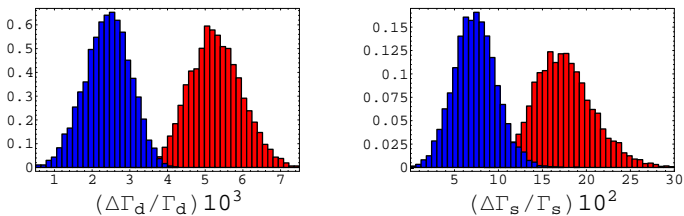}
\vspace*{-0.4cm}
\caption{\label{fig:plot1} Theoretical distributions for $B_d$ and
$B_s$ width differences, at the LO (light/red) and NLO (dark/blue). The
  distributions are those obtained in the (old) basis $\{{\cal
  O}^q_1,{\cal O}^q_2\}$.}
\end{figure}

We note that, after the recent Tevatron measurements of $\Delta M_s$~\cite{CDFD0},
both $\Delta \Gamma_d/\Gamma_d$ and $\Delta \Gamma_s/\Gamma_s$ can be
theoretically obtained within the SM as
\be
\frac{\Delta \Gamma_q}{\Gamma_q} = -
\text{Re}\left(\frac{\Gamma^q_{12}}{M^q_{12}}\right) (\Delta M_q)^\text{exp.} 
\tau(B_q)\,,
\ee
avoiding the quadratic dependence on the decay constants $f_{B_q}$, whose
lattice determinations have still an accuracy of about $15\%$~\cite{Hashimoto:2004hn}.
In spite of that, the theoretical predictions in Eq.~(\ref{eq:dg_noi}) present 
an uncertainty of $\sim 40 \%$, mainly due to strong cancellations coming from 
NLO and $\mathcal{O}(1/m_b^4)$ contributions.
The NLO corrections, indeed, decrease the values of both $\Delta
\Gamma_d/\Gamma_d$ and $\Delta \Gamma_s/\Gamma_s$ by about a factor two with
respect to the LO predictions, as shown in Fig.~\ref{fig:plot1}.
In addition, the $\mathcal{O}(1/m_b^4)$ corrections result of
comparable size relative to the leading $\mathcal{O}(1/m_b^3)$ contributions
and further decrease the width differences.
The theoretical uncertainties in Eq.~(\ref{eq:dg_noi}) turn
out to be dominated by the VSA used for two $\mathcal{O}(1/m_b^4)$ matrix 
elements, followed by the residual NNLO dependence on the renormalization 
scale.

\begin{figure}
\includegraphics[width=0.25\textwidth]{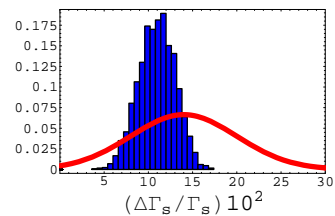}
\vspace*{-0.4cm}
\caption{\label{fig:DGsnew} Theoretical (histogram) vs experimental (solid
  line)  distribution for $\Delta \Gamma_s/\Gamma_s$. The theoretical
  distribution represents the weighted average between the predictions obtained in the
  old and new bases.}
\end{figure}

It was recently observed~\cite{Lenz} that the cancellations
occurring at NLO and $\mathcal{O}(1/m_b^4)$ are
significantly  reduced by expressing the width
differences in the operator basis 
\bea
{\cal O}^q_1 &=& \bar b_i \gamma^\mu (1 - \gamma_5)
  q_i\,\bar b_j \gamma^\mu (1 - \gamma_5) q_j\,,\nn\\ 
{\cal O}^q_3 &=& \bar b_i (1 - \gamma_5)
  q_j\,\bar b_j (1 - \gamma_5) q_i\,,
\eea
instead of the basis $\{{\cal O}^q_1,{\cal O}^q_2\}$ in
Eq.~(\ref{eq:oldbasis}). In the new basis, both NLO and $\mathcal{O}(1/m_b^4)$ 
turn out to be smaller 
and have a reduced impact on the final uncertainties.
Moreover, the Wilson coefficient of the operator ${\cal O}^q_1$ in the new
basis is about ten times larger than in the old basis.
This is a very welcome feature as the ${\cal O}^q_1$ contribution is
the cleanest one, due to the cancellation of the corresponding $B$-parameter in
the ratio $\Gamma_{12}^q/M_{12}^q$.
As a consequence, the theoretical predictions for the width differences that in the new basis read
\be
\Delta \Gamma_{d}/\Gamma_{d} = (4.1 \pm 0.5)\cdot 10^{-3}\,,\quad
\Delta \Gamma_{s}/\Gamma_{s} = (13 \pm 2)\cdot 10^{-2} \,,
\label{eq:dg_new}
\ee
present relative uncertainties reduced by more than a factor two with respect
to the old basis (see Eq.~(\ref{eq:dg_noi})).
One has to note, however, that the significant shifts of central values
due to the change of basis signal important (unknown)
$\mathcal{O}(\alpha_s^2)$ and $\mathcal{O}(\alpha_s/m_b^4)$ corrections.
Therefore, we quote as updated
theoretical predictions the weighted averages of the results obtained in the
two bases, and include in the uncertainty the effects signaled by the shifts
\be
\Delta \Gamma_{d}/\Gamma_{d} = (3.6 \pm 1.0)\cdot 10^{-3}\,,\quad
\Delta \Gamma_{s}/\Gamma_{s} = (11 \pm 4)\cdot 10^{-2}\,.
\label{eq:dg_upd}
\ee
They result to be in good agreement with experimental data, as shown for
$\Delta \Gamma_s/\Gamma_s$ in Fig.~\ref{fig:DGsnew}.

\section{Semileptonic CP-Asymmetries}
\label{sec:3}

The semileptonic CP-asymmetries $A_{SL}^q$ ($q=d,s$) that
describe CP-violation due to mixing, are related to $M_{12}^q$ and 
$\Gamma_{12}^q$, through Eq.~(\ref{eq:dgammared}).
The theoretical predictions of $A_{SL}^q$ are based on the same
perturbative and non-perturbative calculations discussed in Sec.~\ref{sec:2}, 
while the $V_{CKM}$ contributions are different from those 
in  $\Delta \Gamma_q /\Gamma_q$.

\begin{figure}
\includegraphics[width=0.5\textwidth]{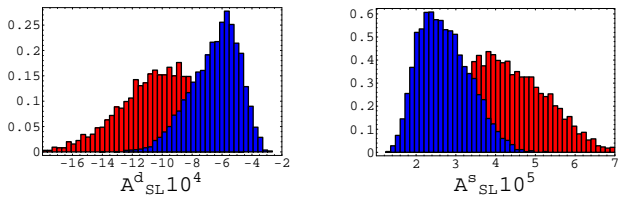}
\vspace*{-0.8cm}
\caption{\label{fig:plot3}
  Theoretical distributions for $B_d$ and $B_s$ semileptonic CP-asymmetries,
  at the LO (light/red) and NLO (dark/blue).}
\end{figure}

The theoretical predictions read~\cite{NOIwip}
\be
\label{eq:nlores3}
A_{SL}^d=-(6.4 \pm 1.6)\cdot 10^{-4}\,,\quad A_{SL}^s=(2.7 \pm 0.6)\cdot 10^{-5}\,. 
\ee
The corresponding theoretical distributions are shown in Fig.~\ref{fig:plot3},
with an evident effect of NLO corrections.

First, still very uncertain, measurements are now
available~\cite{BBpage,D0site} 
\bea  
A_{SL}^d &=& -(30 \pm 78)\cdot 10^{-4}\,,\nn\\
A_{SL}^s &=& (2450 \pm 1930 \pm 350)\cdot 10^{-5}\,. 
\eea
Improved measurements are certainly needed for a significant comparison.
On the theoretical side, $\mathcal{O}(\alpha_s^2)$ and
$\mathcal{O}(\alpha_s/m_b^4)$ contributions are expected to be small
in this case, since the change of basis that had an important impact for width
differences has practically no effect here.

\vspace*{-0.5cm}
\begin{acknowledgments}
\vspace*{-0.3cm}
It is a pleasure to thank my collaborators M.~Ciuchini, E.~Franco, V.~Lubicz
and F.~Mescia, and D.~Becirevic and A.~Lenz for useful discussions.
I warmly thank the CKM2006 organizers for the wonderful
hospitality in Nagoya. 
Partially Supported by the Cluster of Excellence: Origin and Structure of the 
Universe. 
\end{acknowledgments}



\begin{thebibliography}{99}

\bibitem{Buras:1990fn}
  A.~J.~Buras, M.~Jamin and P.~H.~Weisz,
  Nucl.\ Phys.\  B {\bf 347} (1990) 491.

\bibitem{Grossman:1996er}
  Y.~Grossman,
  Phys.\ Lett.\  B {\bf 380} (1996) 99
  [hep-ph/9603244].

\bibitem{Dunietz:2000cr}
  I.~Dunietz, R.~Fleischer and U.~Nierste,
  Phys.\ Rev.\  D {\bf 63} (2001) 114015
  [hep-ph/0012219].

\bibitem{Laplace:2002ik}
  S.~Laplace, Z.~Ligeti, Y.~Nir and G.~Perez,
  Phys.\ Rev.\  D {\bf 65} (2002) 094040
  [hep-ph/0202010].

\bibitem{UTfit}
  M.~Bona {\it et al.}  [UTfit Collaboration],
  JHEP {\bf 0603} (2006) 080
  [hep-ph/0509219]. 
M.~Bona {\it et al.}  [UTfit Collaboration]Phys.\ Rev.\ Lett.\  {\bf 97}
  (2006) 151803 [hep-ph/0605213].
 http://utfit.roma1.infn.it.

\bibitem{BBGT}
  M.~Blanke, A.~J.~Buras, D.~Guadagnoli and C.~Tarantino,
  JHEP {\bf 0610} (2006) 003
  [hep-ph/0604057].

\bibitem{Ligeti:2006pm}
  Z.~Ligeti, M.~Papucci and G.~Perez,
  Phys.\ Rev.\ Lett.\  {\bf 97} (2006) 101801
  [hep-ph/0604112].


\bibitem{ope}
  J.~Chay, H.~Georgi and B.~Grinstein,
  Phys.\ Lett.\  B {\bf 247} (1990) 399.

\bibitem{BBpage}
The Heavy Flavor Averaging Group (HFAG),   
http://www.slac.stanford.edu/xorg/hfag/.


\bibitem{Bigi:1992su}
  I.~I.~Y.~Bigi, N.~G.~Uraltsev and A.~I.~Vainshtein,
  Phys.\ Lett.\  B {\bf 293} (1992) 430
  [Erratum-ibid.\  B {\bf 297} (1993) 477]
  [hep-ph/9207214].

\bibitem{NS}
M.~Neubert and C.T.~Sachrajda,
Nucl.\ Phys.\ B {\textbf 483} (1997) 339
[hep-ph/9603202].

\bibitem{Ciuchini:2001vx}
  M.~Ciuchini, E.~Franco, V.~Lubicz and F.~Mescia,
  Nucl.\ Phys.\  B {\bf 625} (2002) 211
  [hep-ph/0110375].

\bibitem{BBB}
  M.~Beneke {\it et al.},
  Nucl.\ Phys.\  B {\bf 639} (2002) 389
  [hep-ph/0202106].

\bibitem{NOI}
  E.~Franco, V.~Lubicz, F.~Mescia and C.~Tarantino,
  Nucl.\ Phys.\  B {\bf 633} (2002) 212
  [hep-ph/0203089].


\bibitem{Becirevic:2001fy}
D.~Becirevic,
hep-ph/0110124.

\bibitem{DiPierro98}
M.~Di Pierro and C.T.~Sachrajda  [UKQCD Collaboration],
Nucl.\ Phys.\ B {\textbf 534} (1998) 373
[hep-lat/9805028].


\bibitem{DiPierro:1998cj}
M.~Di Pierro and C.~T.~Sachrajda  [UKQCD collaboration],
Nucl.\ Phys.\ Proc.\ Suppl.\  {\textbf 73} (1999) 384
[hep-lat/9809083].

\bibitem{DiPierro:1999tb}
  M.~Di Pierro, C.~T.~Sachrajda and C.~Michael  [UKQCD collaboration],
  Phys.\ Lett.\  B {\bf 468} (1999) 143
  [hep-lat/9906031].

\bibitem{Gabbiani:2003pq}
  F.~Gabbiani, A.~I.~Onishchenko and A.~A.~Petrov,
  Phys.\ Rev.\  D {\bf 68} (2003) 114006
  [hep-ph/0303235];
  Phys.\ Rev.\  D {\bf 70} (2004) 094031
  [hep-ph/0407004].

\bibitem{Tarantino}
  C.~Tarantino,
  Eur.\ Phys.\ J.\  C {\bf 33} (2004) S895
  [hep-ph/0310241];
  Nucl.\ Phys.\ Proc.\ Suppl.\  {\bf 156} (2006) 33
  [hep-ph/0508309].


\bibitem{lambdaCDF}
  A.~Abulencia  [CDF Collaboration],
  hep-ex/0609021.

\bibitem{lambdaD0}
[D0 Collaboration], D0note 5263-Conf,\\
http://www-d0.fnal.gov/.

\bibitem{BBlargh}
  M.~Beneke  {\it et al.},
  Phys.\ Lett.\  B {\bf 459} (1999) 631
  [hep-ph/9808385].

\bibitem{NOIwip}
  M.~Ciuchini {\it et al.},
  JHEP {\bf 0308} (2003) 031
  [hep-ph/0308029].

\bibitem{Beneke:2003az}
  M.~Beneke, G.~Buchalla, A.~Lenz and U.~Nierste,
  Phys.\ Lett.\  B {\bf 576} (2003) 173
  [hep-ph/0307344].

\bibitem{seven}
  M.~Beneke, G.~Buchalla and I.~Dunietz,
  Phys.\ Rev.\  D {\bf 54} (1996) 4419
  [hep-ph/9605259].

\bibitem{Gimenez:2000jj}
V.~Gimenez and J.~Reyes,
Nucl.\ Phys.\ Proc.\ Suppl.\  {\textbf 94} (2001) 350
[hep-lat/0010048].

\bibitem{Hashimoto:2000eh}
  S.~Hashimoto {\it et al.},
  Phys.\ Rev.\  D {\bf 62} (2000) 114502
  [hep-lat/0004022].

\bibitem{Aoki:2002bh}
  S.~Aoki {\it et al.}  [JLQCD Collaboration],
  Phys.\ Rev.\  D {\bf 67} (2003) 014506
  [hep-lat/0208038].

\bibitem{Becirevic:2000sj}
  D.~Becirevic {\it et al.},
  Eur.\ Phys.\ J.\  C {\bf 18} (2000) 157
  [hep-ph/0006135].

\bibitem{Lellouch:2000tw}
L.~Lellouch and C.~J.~Lin  [UKQCD Collaboration],
Phys.\ Rev.\ D {\textbf 64} (2001) 094501
[hep-ph/0011086].

\bibitem{damir}
  D.~Becirevic {\it et al.},
  JHEP {\bf 0204} (2002) 025
  [hep-lat/0110091].

\bibitem{Yamada:2001xp}
  N.~Yamada {\it et al.}  [JLQCD Collaboration],
  Nucl.\ Phys.\ Proc.\ Suppl.\  {\bf 106} (2002) 397
  [hep-lat/0110087].

\bibitem{Aoki:2003xb}
  S.~Aoki {\it et al.}  [JLQCD Collaboration],
  Phys.\ Rev.\ Lett.\  {\bf 91} (2003) 212001
  [hep-ph/0307039].

\bibitem{staggered}
  E.~Dalgic {\it et al.},
  hep-lat/0610104.


\bibitem{pivovarov}
A.~A.~Pivovarov, private communication.

\bibitem{FPCP}
  R.~Van Kooten,
{\it Proceedings of 4th Flavor Physics and CP Violation Conference (FPCP 2006), Vancouver, Canada, 9-12 Apr 2006, pp
031}
  [hep-ex/0606005].

\bibitem{CDFD0}
  A.~Abulencia {\it et al.}  [CDF Collaboration],
  hep-ex/0609040.
  D.~Lucchesi  [CDF and D0 Collaborations],
FERMILAB-CONF-06-262-E.

\bibitem{Hashimoto:2004hn}
  S.~Hashimoto,
  Int.\ J.\ Mod.\ Phys.\  A {\bf 20} (2005) 5133
  [hep-ph/0411126].

\bibitem{Lenz}
  A.~Lenz and U.~Nierste,
  hep-ph/0612167.

\bibitem{D0site}
[D0 Collaboration], D0note 5143-Conf,\\
http://www-d0.fnal.gov/.

\end{thebibliography}
\end{document}